# A Unified Representation and Transformation of Electromagnetic Configurations Based on Generalized Hertz Potentials

**Ting Yi**

Independent Researcher, Irvine, California 92617, USA

E-mail: tingyi.physics@gmail.com

https://orcid.org/0000-0002-2164-5209

## TOC

# A Unified Representation and Transformation of Electromagnetic Configurations Based on Generalized Hertz Potentials

**Ting Yi (蚁艇)**

Independent Researcher, Irvine, California 92617, USA

E-mail: tingyi.physics@gmail.com

https://orcid.org/0000-0002-2164-5209

## Abstract

We present a unified framework that fully represents electromagnetic potentials, fields, and sources in vacuum, based on a generalization of the classical Hertz-potential formalism. In this construction, $\varphi$, $\boldsymbol{A}$, $\boldsymbol{E}$, $\boldsymbol{B}$, $\rho$, and $\boldsymbol{J}$ are systematically derived from a single vector wavefield $\boldsymbol{\Gamma}(\boldsymbol{x}, t)$ (called the Γ-potential), which is structurally aligned with the classical electric Hertz potential but has a broader scope. A surjective mapping is established from such wavefields to all electromagnetic configurations in vacuum (that are sufficiently regular). This mapping induces a well-defined algebraic correspondence between the solution space of Maxwell's equations and the linear space of $C_t^3 C_x^3$ vector wavefields (modulo the relevant symmetries), thereby enabling a framework for structural analysis of electromagnetic fields via their associated wavefields. Gauge freedom and the Lorenz gauge are naturally preserved; charge conservation and Maxwell's equations are inherently encoded in this representation.

Building on this framework, we also introduce a solution-generating transformation that is naturally enabled by the linearity of the representation. This transformation, termed the Γ-transformation, provides a systematic method to derive new, physically distinct electromagnetic configurations from existing ones. It generally leads to changes in potentials, fields, and their underlying sources. Traditional gauge transformations, which alter potentials while preserving fields and sources, can be considered a special, restricted case. An illustrative example is also provided to demonstrate possible applications of the framework.

**Keywords:** Maxwell Equations, Gamma Potential, Gamma Transformation, Electromagnetic Solution Space, Hertz Potentials, Inverse-Source Construction, Retarded Potentials

## 1. Introduction

Mathematically, the physical phenomena of electromagnetism can be described by six quantities as functions of spacetime: charge density $\rho$, current density $\boldsymbol{J}$, electric field $\boldsymbol{E}$, magnetic field $\boldsymbol{B}$, scalar potential $\varphi$, and vector potential $\boldsymbol{A}$.

These six quantities can be naturally grouped into three conceptual pairs. $\rho$ and $\boldsymbol{J}$ form the "source pair", representing the electric charge distribution and its flow. The second pair, containing $\boldsymbol{E}$ and $\boldsymbol{B}$, is the "field pair", representing the physical electromagnetic fields, measurable through their effects. The third is the "potential pair"—$\varphi$ and $\boldsymbol{A}$—often regarded as mathematical constructs for expressing the fields, though their possible physical significance is suggested by the Aharonov–Bohm effect [1]. Due to gauge freedom, the potentials are not uniquely defined; however, under suitable conditions they are determined by the fields up to a gauge transformation.

These three pairs describe electromagnetic phenomena at three distinct but interconnected levels. They are not independent. Within and across these pairs, comprehensive mathematical relationships exist and are explicitly described by a set of equations [2, 3]. Charges and currents act as the sources of the fields and, in principle, determine the full electromagnetic configuration. Fields describe the observable effects that form the basis of measurement and practical applications. Potentials, beyond their possible physical significance, provide a mathematical representation of the fields.

Such interconnection and interdependence are not merely qualitative. The differential relations that tie these six quantities are mutually constraining. This suggests that a more compact underlying structure may exist, enabling a more unified organization of the full electromagnetic configuration.

In this regard, the Hertz potential [3–5] may already provide a meaningful clue. Although it is often introduced as a technical tool for solving polarization and radiation problems in homogeneous media, the Hertz potential can generate both scalar and vector potentials—and hence the electric and magnetic fields—from a single vector function. This construction hints at the possibility of extending it to represent the full electromagnetic configuration, at a deeper structural level.

However, this structural perspective on the Hertz formalism remains largely underexplored in the literature. The present work is motivated by this observation, and seeks to reinterpret and extend the Hertz framework into a general representation capable of describing all electromagnetic configurations in vacuum—including those involving arbitrary sources—with full structural clarity. This representation, which we call the **Γ-representation**, forms the theoretical foundation of this work.

As is well known, the interrelations among the electromagnetic quantities are governed by several fundamental equations. For completeness and consistency of notation and format, we list them below.

Within the source pair, charge conservation is expressed by the continuity equation:

$$\frac{\partial \rho}{\partial t} + \nabla \cdot \boldsymbol{J} = 0. \tag{1}$$

This indicates that electric charge is a conserved quantity, and current density represents its motion.

For the field pair, the dynamics and coupling to sources are described by Maxwell's equations, which in SI units (with $c = 1/\sqrt{\varepsilon_0 \mu_0}$ ) can be written as:

$$\nabla \cdot \boldsymbol{E} = \frac{\rho}{\varepsilon_0}, \tag{2}$$

$$\nabla \cdot \boldsymbol{B} = 0, \tag{3}$$

$$\nabla \times \boldsymbol{E} = -\frac{\partial \boldsymbol{B}}{\partial t}, \tag{4}$$

$$\nabla \times \boldsymbol{B} = \frac{1}{c^2}\frac{\partial \boldsymbol{E}}{\partial t} + \mu_0 \boldsymbol{J}. \tag{5}$$

These equations are internally consistent and compatible with the continuity equation, illustrating the redundancy in the description.

Within the potential pair, a gauge condition imposes a constraint between $\varphi$ and $\boldsymbol{A}$ without altering the physical fields. In this paper, the Lorenz gauge condition is adopted:

$$\frac{1}{c^2}\frac{\partial \varphi}{\partial t} + \nabla \cdot \boldsymbol{A} = 0. \tag{6}$$

The fields can be expressed in terms of the potentials via:

$$\boldsymbol{B} = \nabla \times \boldsymbol{A}, \tag{7}$$

$$\boldsymbol{E} = -\frac{\partial \boldsymbol{A}}{\partial t} - \nabla \varphi. \tag{8}$$

Finally, under the Lorenz gauge, the relation between potentials and sources satisfies the following wave equations (which can be derived from Eqs. (2)–(8)):

$$\frac{1}{c^2}\frac{\partial^2 \varphi}{\partial t^2} - \nabla^2 \varphi = \frac{\rho}{\varepsilon_0}, \tag{9}$$

$$\frac{1}{c^2}\frac{\partial^2 \boldsymbol{A}}{\partial t^2} - \nabla^2 \boldsymbol{A} = \mu_0 \boldsymbol{J}. \tag{10}$$

In this paper, we refer to the collection of electromagnetic potentials, fields, and sources as an **electromagnetic configuration** (or **EM configuration** in short). This term encompasses the scalar and vector potentials $\varphi$ and $\boldsymbol{A}$, the electric and magnetic fields $\boldsymbol{E}$ and $\boldsymbol{B}$, and the associated charge-current sources $\rho$ and $\boldsymbol{J}$.

Unless otherwise specified, potentials are assumed to satisfy the Lorenz gauge condition in this paper. In contexts involving the solution space of Maxwell's equations, the term **EM solution** will also be used interchangeably with EM configuration, when no ambiguity arises.

To facilitate notation, we compactly denote an EM configuration (with its six quantities collectively satisfying Eqs. (1)–(8)) as

$$[\mathrm{EM}] \triangleq [\varphi, \boldsymbol{A}, \boldsymbol{E}, \boldsymbol{B}, \rho, \boldsymbol{J}].$$

Accordingly, we write $[\mathrm{EM}_1]$, $[\mathrm{EM}_2]$,… to label distinct configurations, and write $[\mathrm{EM}(\boldsymbol{\Gamma})] \triangleq [\varphi_\Gamma, \boldsymbol{A}_\Gamma, \boldsymbol{E}_\Gamma, \boldsymbol{B}_\Gamma, \rho_\Gamma, \boldsymbol{J}_\Gamma]$ to denote the configuration associated with a given vector $\boldsymbol{\Gamma}$.

The structure of this paper is as follows: Section 2 introduces the main construction and presents the general theorem and its converse. Section 3 discusses the symmetries of the EM solutions. Section 4 is dedicated to the Γ-transformation. Section 5 further explores structure of the solution space of Maxwell's equations. Section 6 applies the proposed representation to electromagnetic fields in media and analyzes its relation to Hertz potentials. Section 7 presents an additional illustrative application example. Section 8 offers a summary and outlook on further research.

## 2. Γ-Representation: The Main Theorems and Proofs

We now introduce the main mathematical structure of the proposed representation—a general construction that derives electromagnetic potentials, fields, and sources from a single vector wavefield.

**Theorem 1 (Γ-Representation Theorem).**

Let $\boldsymbol{\Gamma}(\boldsymbol{x},t) \in C_t^3 C_x^3(\mathbb{R} \times \mathbb{R}^3; \mathbb{R}^3)$ be a vector field satisfying the vector wave equation:

$$\Box\boldsymbol{\Gamma}(\boldsymbol{x},t) \triangleq \frac{1}{c^2}\frac{\partial^2 \boldsymbol{\Gamma}(\boldsymbol{x},t)}{\partial t^2} - \nabla^2 \boldsymbol{\Gamma}(\boldsymbol{x},t) = \boldsymbol{G}(\boldsymbol{x},t). \tag{11}$$

Define:

$$\varphi_\Gamma = -\nabla \cdot \boldsymbol{\Gamma}, \tag{12}$$

$$\boldsymbol{A}_\Gamma = \frac{1}{c^2}\frac{\partial \boldsymbol{\Gamma}}{\partial t}, \tag{13}$$

$$\boldsymbol{E}_\Gamma = -\frac{1}{c^2}\frac{\partial^2 \boldsymbol{\Gamma}}{\partial t^2} + \nabla(\nabla \cdot \boldsymbol{\Gamma}), \tag{14}$$

or an alternative form (shown in the proof): $\boldsymbol{E}_\Gamma = \nabla \times (\nabla \times \boldsymbol{\Gamma}) - \boldsymbol{G}$, (14a)

$$\boldsymbol{B}_\Gamma = \frac{1}{c^2}\frac{\partial}{\partial t}(\nabla \times \boldsymbol{\Gamma}), \tag{15}$$

$$\rho_\Gamma = -\varepsilon_0 \nabla \cdot \boldsymbol{G}, \tag{16}$$

$$\boldsymbol{J}_\Gamma = \varepsilon_0 \frac{\partial \boldsymbol{G}}{\partial t}. \tag{17}$$

Then:

1. $\varphi_\Gamma$ and $\boldsymbol{A}_\Gamma$, regarded as the scalar and vector potentials, satisfy the Lorenz gauge condition.

2. $\boldsymbol{E}_\Gamma$ and $\boldsymbol{B}_\Gamma$, considered as the electric and magnetic fields, are related to $\varphi_\Gamma$ and $\boldsymbol{A}_\Gamma$ via the standard field–potential relations of classical electrodynamics (i.e., Eqs. (7) and (8)).

3. $\rho_\Gamma$ and $\boldsymbol{J}_\Gamma$, interpreted as the charge and current densities, together with $\boldsymbol{E}_\Gamma$ and $\boldsymbol{B}_\Gamma$, satisfy Maxwell's equations in vacuum, along with the continuity equation for charge conservation.

**Proof**.

The assumption $\boldsymbol{\Gamma} \in C_t^3 C_x^3$ ensures that all derivatives appearing in Eqs. (11)–(17) and in Maxwell's equations exist as classical derivatives and are continuous; mixed derivatives commute.

Eq. (6), the Lorenz gauge condition, follows from Eqs. (12) and (13):

$$\frac{1}{c^2}\frac{\partial \varphi_\Gamma}{\partial t} + \nabla \cdot \boldsymbol{A}_\Gamma = -\frac{1}{c^2}\frac{\partial}{\partial t}(\nabla \cdot \boldsymbol{\Gamma}) + \nabla \cdot \left(\frac{1}{c^2}\frac{\partial \boldsymbol{\Gamma}}{\partial t}\right) = 0.$$

The two expressions for the electric field, Eqs. (14) and (14a), are equivalent. This can be confirmed by applying the vector identity $\nabla^2\boldsymbol{\Gamma} = \nabla(\nabla\cdot\boldsymbol{\Gamma}) - \nabla\times(\nabla\times\boldsymbol{\Gamma})$ to Eq. (11) and reorganizing terms.

Eq. (7), which relates the vector potential to the magnetic field, follows directly from Eqs. (13) and (15).

Eq. (8), which expresses the electric field in terms of the potentials, follows by inserting Eqs. (12) and (13) into Eq. (14):

$$\boldsymbol{E}_\Gamma = -\frac{1}{c^2}\frac{\partial}{\partial t}\left(\frac{\partial\boldsymbol{\Gamma}}{\partial t}\right) + \nabla(\nabla\cdot\boldsymbol{\Gamma}) = -\frac{\partial\boldsymbol{A}_\Gamma}{\partial t} - \nabla\varphi_\Gamma.$$

To verify Maxwell's equations, Eqs. (2)–(5):

- Eq. (2) (Gauss's law) follows by taking the divergence of Eq. (14a), using Eq. (16), and noting that the divergence of a curl vanishes:

$$\nabla\cdot\boldsymbol{E}_\Gamma = \nabla\cdot\big(\nabla\times(\nabla\times\boldsymbol{\Gamma})\big) - \nabla\cdot\boldsymbol{G} = \frac{\rho_\Gamma}{\varepsilon_0}.$$

- Eq. (3) (Gauss's law for magnetism) holds since $\boldsymbol{B}_\Gamma = \nabla\times\boldsymbol{A}_\Gamma$ (Eq. (7)) and the divergence of any curl is zero.

- Eq. (4) (Faraday's law of induction) follows by taking the curl of Eq. (14):

$$\nabla\times\boldsymbol{E}_\Gamma = -\frac{1}{c^2}\frac{\partial^2(\nabla\times\boldsymbol{\Gamma})}{\partial t^2} + \nabla\times\big(\nabla(\nabla\cdot\boldsymbol{\Gamma})\big) = -\frac{\partial}{\partial t}\left(\frac{1}{c^2}\frac{\partial}{\partial t}(\nabla\times\boldsymbol{\Gamma})\right) = -\frac{\partial\boldsymbol{B}_\Gamma}{\partial t}.$$

- Eq. (5) (Ampère–Maxwell law) follows from rewriting Eq. (14a) as $\nabla\times(\nabla\times\boldsymbol{\Gamma}) = \boldsymbol{E}_\Gamma + \boldsymbol{G}$ and substituting it into the curl of Eq. (15), and then using Eq. (17):

$$\nabla\times\boldsymbol{B}_\Gamma = \frac{1}{c^2}\frac{\partial}{\partial t}\big(\nabla\times(\nabla\times\boldsymbol{\Gamma})\big) = \frac{1}{c^2}\frac{\partial\boldsymbol{E}_\Gamma}{\partial t} + \frac{1}{c^2}\frac{\partial\boldsymbol{G}}{\partial t} = \frac{1}{c^2}\frac{\partial\boldsymbol{E}_\Gamma}{\partial t} + \mu_0\boldsymbol{J}_\Gamma.$$

Finally, Eq. (1), the continuity equation, follows from Eqs. (16)–(17):

$$\frac{\partial\rho_{\boldsymbol{\Gamma}}}{\partial t} + \nabla\cdot\boldsymbol{J}_\Gamma = -\varepsilon_0\frac{\partial(\nabla\cdot\boldsymbol{G})}{\partial t} + \varepsilon_0\nabla\cdot\left(\frac{\partial\boldsymbol{G}}{\partial t}\right) = 0.$$

Q.E.D.

For Maxwell's equations and other identities in Eqs. (1)–(10) to hold pointwise, a minimal regularity threshold is required. In particular, it is necessary (and sufficient) that $\rho \in C_t^1C_x^0$ and $\boldsymbol{J} \in C_t^0C_x^1$ for the continuity equation (Eq. (1)) to hold in the strong (pointwise) sense. For Eqs. (9)–(10) with source $(\rho,\boldsymbol{J})$ to hold pointwise, it is required that $\boldsymbol{A} \in C_t^2C_x^3$ and $\phi \in C_t^3C_x^2$. These, in turn, imply $\boldsymbol{B} \in C_t^2C_x^2$ via Eq. (7) and $\boldsymbol{E} \in C_t^1C_x^1$ via Eq. (8).

We refer to vacuum configurations meeting these regularity requirements as **regular electromagnetic configurations**. Cases with spatial or temporal jumps/impulses (e.g., line/sheet charges or time-impulse currents) fall below this continuity threshold; some identities then hold only in the distributional (weak) sense. These "non-regular" configurations are excluded from the discussion in this paper.

The assumption in Theorem 1, $\boldsymbol{\Gamma} \in C_t^3C_x^3$, is sufficient and essentially minimal (within this framework) to ensure that the induced potentials, fields, and sources form a regular electromagnetic configuration.

Applying the notation introduced in Section 1, Theorem 1 essentially states that a vector wavefield $\boldsymbol{\Gamma}$, together with its associated source field $\boldsymbol{G}$, gives rise to an EM configuration $[\mathrm{EM}(\boldsymbol{\Gamma})] = [\varphi_\Gamma, \boldsymbol{A}_\Gamma, \boldsymbol{E}_\Gamma, \boldsymbol{B}_\Gamma, \rho_\Gamma, \boldsymbol{J}_\Gamma]$ through Eqs. (12)–(17). In other words, the full electromagnetic configuration—including the fields, potentials, and sources—along with Maxwell's equations and other interrelations, is fully encoded in the single wavefield $\boldsymbol{\Gamma}$.

Such configurations will be referred to as **Γ-representations**, where the wavefield $\boldsymbol{\Gamma}$ is called the **Γ-potential**, and the corresponding source field $\boldsymbol{G}$ is termed the **Γ-source**.

According to Theorem 1, a spacetime vector field $\boldsymbol{\Gamma}$ qualifies as a valid Γ-potential if it is of class $C_t^3 C_x^3$, ensuring that $\Box\boldsymbol{\Gamma}$ and its first derivatives are well defined and the induced mappings to ($\varphi$, $\boldsymbol{A}$, $\boldsymbol{E}$, $\boldsymbol{B}$, $\rho$, $\boldsymbol{J}$) are classical (i.e., valid pointwise).

This mapping, $\boldsymbol{\Gamma} \longmapsto [\mathrm{EM}(\boldsymbol{\Gamma})]$, however, is not injective: if two Γ-potentials differ by a constant vector field, they yield the same EM configuration. This nontrivial degree of freedom will be further discussed in Section 3.

A natural question then arises: is this mapping surjective—that is, can every regular electromagnetic configuration in vacuum be realized as a Γ-representation? The answer is affirmative. Every regular EM configuration (i.e., satisfying the continuity threshold stated above) admits at least one corresponding Γ-potential, though not necessarily unique. This is formalized in the following theorem.

**Theorem 2 (Converse Γ-Representation Theorem).**

Let $[\mathrm{EM}] = [\varphi, \boldsymbol{A}, \boldsymbol{E}, \boldsymbol{B}, \rho, \boldsymbol{J}]$ be a regular vacuum electromagnetic configuration. Then there exists a spacetime vector field $\boldsymbol{\Gamma}$ that serves as the Γ-potential of [EM].

In other words, Theorem 2 asserts that every vacuum solution of Maxwell's equations that meets the continuity threshold (i.e., the "regular" class) can be fully represented by a single $C_t^3 C_x^3$ vector wavefield (as the Γ-potential). To prove this, we directly construct such a Γ-potential from an arbitrary regular EM configuration $[\varphi, \boldsymbol{A}, \boldsymbol{E}, \boldsymbol{B}, \rho, \boldsymbol{J}]$.

**Proof**.

As scalar and vector potentials of a regular EM configuration, we have $\varphi \in C_t^3 C_x^2$ and $\boldsymbol{A} \in C_t^2 C_x^3$. Fix a reference time $t_0$ and choose a spatial vector field $\boldsymbol{F}_{t0}(\boldsymbol{x}) \in C_x^3$ such that

$$\nabla \cdot \boldsymbol{F}_{t0}(\boldsymbol{x}) = -\varphi(\boldsymbol{x}, t_0).$$

Define the spacetime vector field:

$$\boldsymbol{\Gamma}(\boldsymbol{x}, t) = \boldsymbol{F}_{t0}(\boldsymbol{x}) + \int_{t_0}^{t} c^2 \boldsymbol{A}(\boldsymbol{x}, \tau) d\tau. \tag{18}$$

Then $\boldsymbol{\Gamma} \in C_t^3 C_x^3$, so it qualifies as a Γ-potential. Such regularity also guarantees the existence of a vector field $\boldsymbol{G} \in C_t^1 C_x^1$ satisfying the wave equation $\Box\boldsymbol{\Gamma}(\boldsymbol{x}, t) = \boldsymbol{G}$.

We now verify that such a vector field $\boldsymbol{\Gamma}$, as a Γ-potential, reproduces the given EM configuration $[\varphi, \boldsymbol{A}, \boldsymbol{E}, \boldsymbol{B}, \rho, \boldsymbol{J}]$ via Eqs. (12)–(17). In other words, with $\boldsymbol{\Gamma}$ given by Eq. (18), the relations in Eqs. (12)–(17) hold identically and recover the prescribed fields, potentials, and sources in [EM].

Eq. (12) follows by taking the divergence of $\boldsymbol{\Gamma}$ in Eq. (18) and applying the Lorenz gauge condition:

$$\nabla\cdot\boldsymbol{\Gamma} = \nabla\cdot\boldsymbol{F}_{t0} + \int_{t_0}^{t} c^2(\nabla\cdot\boldsymbol{A})d\tau = -\varphi(\boldsymbol{x},t_0) - \int_{t_0}^{t} c^2\left(\frac{1}{\mathrm{c}^2}\frac{\partial\varphi}{\partial t}\right)d\tau = -\varphi(\boldsymbol{x},t).$$

Eq. (13) arises from the time derivative of $\boldsymbol{\Gamma}$ in Eq. (18) and the time-independence of $\boldsymbol{F}_{t0}(\boldsymbol{x})$:

$$\frac{\partial}{\partial t}\boldsymbol{\Gamma}(\boldsymbol{x},t) = \frac{\partial}{\partial t}\boldsymbol{F}_{t0}(\boldsymbol{x}) + \frac{\partial}{\partial t}\int_{t_0}^{t} c^2\boldsymbol{A}(\boldsymbol{x},\tau)d\tau = c^2\boldsymbol{A}(\boldsymbol{x},t),$$

Eq. (15) can be obtained by inserting Eq. (13) into Eq. (7):

$$\boldsymbol{B}(\boldsymbol{x},t) = \nabla\times\boldsymbol{A}(\boldsymbol{x},t) = \nabla\times\left(\frac{1}{c^2}\frac{\partial}{\partial t}\boldsymbol{\Gamma}(\boldsymbol{x},t)\right) = \frac{1}{c^2}\frac{\partial}{\partial t}\big(\nabla\times\boldsymbol{\Gamma}(\boldsymbol{x},t)\big).$$

Eq. (14) is yielded by inserting $\varphi$ (Eq. (12)) and $\boldsymbol{A}$ (Eq. (13)) into $\boldsymbol{E} = -\frac{\partial\boldsymbol{A}}{\partial t} - \nabla\varphi$ (Eq. (8)):

$$\boldsymbol{E} = -\frac{\partial}{\partial t}\left(\frac{1}{c^2}\frac{\partial}{\partial t}\boldsymbol{\Gamma}\right) - \nabla(-\nabla\cdot\boldsymbol{\Gamma}) = -\frac{1}{c^2}\frac{\partial^2}{\partial t^2}\boldsymbol{\Gamma} + \nabla(\nabla\cdot\boldsymbol{\Gamma}).$$

Eq. (14a) is then recovered applying the wave equation $\Box\boldsymbol{\Gamma} = \boldsymbol{G}$ to Eq. (14):

$$\boldsymbol{E} = -\frac{1}{c^2}\frac{\partial^2}{\partial t^2}\boldsymbol{\Gamma} + \nabla(\nabla\cdot\boldsymbol{\Gamma}) = \left(-\frac{1}{c^2}\frac{\partial^2}{\partial t^2}\boldsymbol{\Gamma} + \nabla^2\boldsymbol{\Gamma}\right) + \nabla\times(\nabla\times\boldsymbol{\Gamma}) = -\boldsymbol{G} + \nabla\times(\nabla\times\boldsymbol{\Gamma}).$$

Eq. (16) is a consequence of taking the divergence of Eq. (14a) and applying Eq. (2):

$$\nabla\cdot\boldsymbol{E} = \nabla\cdot\big(\nabla\times(\nabla\times\boldsymbol{\Gamma})\big) - \nabla\cdot\boldsymbol{G} = -\nabla\cdot\boldsymbol{G} = \frac{\rho}{\varepsilon_0}.$$

Finally, for Eq. (17), substituting Eqs. (14) and (15) into Eq. (5) yields:

$$\nabla\times\left(\frac{1}{c^2}\frac{\partial}{\partial t}(\nabla\times\boldsymbol{\Gamma})\right) = \frac{1}{c^2}\frac{\partial}{\partial t}\left(-\frac{1}{c^2}\frac{\partial^2}{\partial t^2}\boldsymbol{\Gamma}(\boldsymbol{x},t) + \nabla(\nabla\cdot\boldsymbol{\Gamma})\right) + \mu_0\boldsymbol{J},$$

which simplifies to:

$$\frac{1}{c^2}\frac{\partial}{\partial t}\big(\nabla\times(\nabla\times\boldsymbol{\Gamma})\big) = \frac{1}{c^2}\frac{\partial}{\partial t}\left(-\frac{1}{c^2}\frac{\partial^2}{\partial t^2}\boldsymbol{\Gamma}(\boldsymbol{x},t) + \nabla(\nabla\cdot\boldsymbol{\Gamma})\right) + \mu_0\boldsymbol{J}.$$

Reorganizing terms and applying the wave equation $\Box\boldsymbol{\Gamma} = \boldsymbol{G}$, we obtain:

$$\frac{1}{c^2}\frac{\partial}{\partial t}\left(\frac{1}{c^2}\frac{\partial^2}{\partial t^2}\boldsymbol{\Gamma}(\boldsymbol{x},t) - \nabla(\nabla\cdot\boldsymbol{\Gamma}) + \nabla\times(\nabla\times\boldsymbol{\Gamma})\right) = \frac{1}{c^2}\frac{\partial}{\partial t}(\boldsymbol{\Gamma}) = \frac{1}{c^2}\frac{\partial\boldsymbol{G}}{\partial t} = \mu_0\boldsymbol{J}.$$

Hence $\boldsymbol{J} = \frac{1}{\mu_0 c^2}\frac{\partial\boldsymbol{G}}{\partial t} = \varepsilon_0\frac{\partial\boldsymbol{G}}{\partial t}$, which is Eq. (17).

Q.E.D.

The above construction of $\boldsymbol{\Gamma}$ from a given electromagnetic configuration further illustrates the role of Γ-potential. As defined in Eq. (18), $\boldsymbol{\Gamma}$ is essentially the time integral of vector potential $\boldsymbol{A}$, with an initial condition $\boldsymbol{F}_{t0}(\boldsymbol{x})$ which is effectively the "spatial integral" or "inverse divergence" of scalar potential $\varphi$ at the initial time $t_0$.

More generally, the significance of the Γ-potential can be understood through Eqs. (12) and (13). Specifically, $\boldsymbol{\Gamma}$ may be viewed either as the preimage of scalar potential $\varphi$ under the divergence operator (Eq. (12)), or as the preimage of vector potential $\boldsymbol{A}$ under the time derivative (Eq. (13)). The Lorenz gauge condition bridges these two preimages, enabling the single function $\boldsymbol{\Gamma}$ to represent both $\varphi$ and $\boldsymbol{A}$, and thereby represent the full electromagnetic configuration—including fields $\boldsymbol{E}$ and $\boldsymbol{B}$, and sources $\rho$ and $\boldsymbol{J}$.

If the three canonical pairs in electrodynamics—namely, the source pair ($\rho$, $\boldsymbol{J}$), the field pair ($\boldsymbol{E}$, $\boldsymbol{B}$), and the potential pair ($\varphi$, $\boldsymbol{A}$)—are understood as describing electromagnetic phenomena at three successive levels, then the Γ-potential may be interpreted as extending this hierarchy by adding a fourth layer beyond the potentials.

As can be seen from its mathematical construction, the concept of Γ-potential bears a strong resemblance to the classical Hertz potentials—particularly the electric Hertz potential [3–5]. At the heart of the Hertz formalism lies a simple yet profound idea: using a single vector function to generate both the scalar and vector potentials, and thereby the complete electromagnetic fields. The Γ-potential inherits this spirit and attempts to generalize it into a symmetric and comprehensive framework that applies to all regular electromagnetic configurations in vacuum, including those involving charge and current sources.

The relationship between the Hertz potentials and the Γ-potential will be further discussed in Section 6.

## 3. Symmetry in Γ-Representation

From the perspective of their interrelations, the four hierarchical levels of electromagnetic quantities—from sources, fields to potentials and the Γ-potential—are connected through successive applications of differential operators. At each level, the quantities can be obtained from those at one level below via operations such as gradient, divergence, curl, or time derivative. Specifically, the sources $\rho$ and $\boldsymbol{J}$, as expressed in Eqs. (2) and (5), can be derived from combinations of divergence, curl, and time derivative of the fields $\boldsymbol{E}$ and $\boldsymbol{B}$; the electric and magnetic fields $\boldsymbol{E}$ and $\boldsymbol{B}$, in turn, can be obtained from the scalar and vector potentials via gradient, curl, and time derivative operations, as shown in Eqs. (7) and (8). Finally, according to Eqs. (12) and (13), the potentials $\varphi$ and $\boldsymbol{A}$ can be derived from the Γ-potential through divergence and time differentiation.

As a natural consequence of this nested differential structure, each level admits certain degrees of freedom—analogous to "integration constants"—that leave all higher-level quantities unchanged. At the potential level, for example, such freedom appears as the conventional gauge symmetry, which transforms scalar and vector potentials while leaving the resulting fields and sources invariant. At the field level, adding electric and magnetic fields of a source-free wavefield to an existing EM solution does not alter the corresponding sources. Within the Γ-representation framework, these structural freedoms across all four levels can be systematically characterized as structural invariances. This offers a unified and deeper expression of symmetry that may reveal the intrinsic mathematical nature of electromagnetic configurations.

The following theorem characterizes these invariances in a systematic and hierarchical manner.

**Theorem 3 (Γ-Representation Invariance Theorem).**

Let $\boldsymbol{\Gamma}_1$ and $\boldsymbol{\Gamma}_2$ be two Γ-potentials corresponding to electromagnetic configurations $[\mathrm{EM}_1] = [\varphi_1, \boldsymbol{A}_1, \boldsymbol{E}_1, \boldsymbol{B}_1, \rho_1, \boldsymbol{J}_1]$ and $[\mathrm{EM}_2] = [\varphi_2, \boldsymbol{A}_2, \boldsymbol{E}_2, \boldsymbol{B}_2, \rho_2, \boldsymbol{J}_2]$, respectively. Define $\boldsymbol{\gamma} = \boldsymbol{\Gamma}_2 - \boldsymbol{\Gamma}_1$. Then:

(a) If $\nabla \cdot \boldsymbol{\gamma} = 0$ and $\frac{\partial \boldsymbol{\gamma}}{\partial t} = 0$, then $[\mathrm{EM}_2] = [\mathrm{EM}_1]$.

(b) If $\Box \boldsymbol{\gamma} = 0$ and $\nabla \times (\nabla \times \boldsymbol{\gamma}) = 0$, and assuming a simply connected domain and standard decay at infinity for $\boldsymbol{B}_2 - \boldsymbol{B}_1$, then $\boldsymbol{E}_2 = \boldsymbol{E}_1$, $\boldsymbol{B}_2 = \boldsymbol{B}_1$, $\rho_2 = \rho_1$, and $\boldsymbol{J}_2 = \boldsymbol{J}_1$; moreover, the potentials $(\varphi_2, \boldsymbol{A}_2)$ and $(\varphi_1, \boldsymbol{A}_1)$ are related by a gauge transformation that preserves the Lorenz gauge condition.

(c) If $\Box \boldsymbol{\gamma} = 0$, then $\rho_2 = \rho_1$ and $\boldsymbol{J}_2 = \boldsymbol{J}_1$.

**Proof.**

As assumed, $\boldsymbol{\Gamma}_2 = \boldsymbol{\Gamma}_1 + \boldsymbol{\gamma}$. Substituting into Eqs. (12) and (13), we obtain:

$$\varphi_2 = \varphi_1 - \boldsymbol{\nabla} \cdot \boldsymbol{\gamma}, \quad \boldsymbol{A}_2 = \boldsymbol{A}_1 + \frac{1}{c^2}\frac{\partial \boldsymbol{\gamma}}{\partial t}. \tag{19}$$

(a) If $\nabla \cdot \boldsymbol{\gamma} = 0$ and $\frac{\partial \boldsymbol{\gamma}}{\partial t} = 0$, then Eq. (19) yields $\varphi_2 = \varphi_1$ and $\boldsymbol{A}_2 = \boldsymbol{A}_1$. Since both scalar and vector potentials are equal, the corresponding fields and sources derived from them are also equal. Thus, $[\mathrm{EM}_2] = [\mathrm{EM}_1]$.

(b) From $\Box \boldsymbol{\gamma} = \frac{1}{c^2}\frac{\partial^2 \boldsymbol{\gamma}}{\partial t^2} - \nabla^2 \boldsymbol{\gamma} = 0$ and $\nabla \times (\nabla \times \boldsymbol{\gamma}) = \nabla(\nabla \cdot \boldsymbol{\gamma}) - \nabla^2 \boldsymbol{\gamma} = 0$, we obtain

$$\frac{1}{c^2}\frac{\partial^2 \boldsymbol{\gamma}}{\partial t^2} = \nabla(\nabla \cdot \boldsymbol{\gamma}). \tag{20}$$

Define a scalar $\chi(\boldsymbol{x}, t)$ by prescribing its time derivative,

$$\frac{\partial \chi}{\partial t} = \nabla \cdot \boldsymbol{\gamma}. \tag{21}$$

Taking the gradient of Eq. (21) and using Eq. (20) gives

$$\nabla \frac{\partial \chi}{\partial t} = \frac{\partial}{\partial t}(\nabla \chi) = \nabla(\nabla \cdot \boldsymbol{\gamma}) = \frac{1}{c^2}\frac{\partial^2 \boldsymbol{\gamma}}{\partial t^2},$$

hence

$$\frac{\partial}{\partial t}\left(\frac{1}{c^2}\frac{\partial \boldsymbol{\gamma}}{\partial t} - \nabla \chi\right) = 0. \tag{22}$$

That is, $\frac{1}{c^2}\frac{\partial \boldsymbol{\gamma}}{\partial t} - \nabla \chi$ is time-independent, and so is its curl:

$$\boldsymbol{K} \triangleq \nabla \times \left(\frac{1}{c^2}\frac{\partial \boldsymbol{\gamma}}{\partial t} - \nabla \chi\right) = \frac{1}{c^2}\frac{\partial}{\partial t}(\nabla \times \boldsymbol{\gamma}). \tag{23}$$

Besides being time-independent, $\boldsymbol{K}$ also satisfies

$$\nabla \cdot \boldsymbol{K} = \frac{1}{c^2}\frac{\partial}{\partial t}\big(\nabla \cdot (\nabla \times \boldsymbol{\gamma})\big) = 0, \quad \nabla \times \boldsymbol{K} = \frac{1}{c^2}\frac{\partial}{\partial t}\big(\nabla \times (\nabla \times \boldsymbol{\gamma})\big) = 0.$$

Substituting these into the vector identity $\Delta\boldsymbol{K} = \nabla(\nabla\cdot\boldsymbol{K}) - \nabla\times(\nabla\times\boldsymbol{K})$ gives $\Delta\boldsymbol{K} = 0$; i.e., $\boldsymbol{K}$ is a harmonic vector field. Moreover, by Eq. (15), we actually have

$$\boldsymbol{B}_2 - \boldsymbol{B}_1 = \frac{1}{c^2}\frac{\partial}{\partial t}(\nabla\times\boldsymbol{\gamma}) = \boldsymbol{K}.$$

The assumed standard decay at infinity for $\boldsymbol{B}_2 - \boldsymbol{B}_1$ implies that $\boldsymbol{K}(\boldsymbol{x}) \to 0$ as $|\boldsymbol{x}| \to \infty$. Liouville's theorem and standard uniqueness for decaying harmonic fields then ensure $\boldsymbol{K} \equiv 0$.

With Eq. (22) and $\boldsymbol{K} \equiv 0$ in Eq. (23), the vector $\frac{1}{c^2}\frac{\partial\boldsymbol{\gamma}}{\partial t} - \nabla\chi$ is time-independent and curl-free. Hence, on a simply connected domain (or in $\mathbb{R}^3$ with decay), we can write

$$\frac{1}{c^2}\frac{\partial\boldsymbol{\gamma}}{\partial t} - \nabla\chi = \nabla f(x). \tag{24}$$

for some scalar $f(x)$. Replacing $\chi$ by $\chi + f(x)$, Eq. (21) still stands and Eq. (24) gives

$$\nabla\chi = \frac{1}{c^2}\frac{\partial\boldsymbol{\gamma}}{\partial t}. \tag{25}$$

Substituting Eq. (25) and (21) into Eq. (19), we find

$$\varphi_2 = \varphi_1 - \frac{\partial\chi}{\partial t}, \quad \boldsymbol{A}_2 = \boldsymbol{A}_1 + \nabla\chi.$$

Thus $(\varphi_2, \boldsymbol{A}_2)$ and $(\varphi_1, \boldsymbol{A}_1)$ are related by a gauge transformation, hence the corresponding fields and sources remain unchanged: $\boldsymbol{E}_2 = \boldsymbol{E}_1$, $\boldsymbol{B}_2 = \boldsymbol{B}_1$, $\rho_2 = \rho_1$, and $\boldsymbol{J}_2 = \boldsymbol{J}_1$.

Finally, since the gauge function $\chi$ satisfies

$$\Box\chi = \frac{1}{c^2}\frac{\partial^2\chi}{\partial t^2} - \nabla^2\chi = \frac{1}{c^2}\frac{\partial}{\partial t}\left(\frac{\partial\chi}{\partial t}\right) - \nabla\cdot(\nabla\chi) = \frac{1}{c^2}\frac{\partial}{\partial t}(\nabla\cdot\boldsymbol{\gamma}) - \nabla\cdot\left(\frac{1}{c^2}\frac{\partial\boldsymbol{\gamma}}{\partial t}\right) = 0,$$

the Lorenz gauge condition is preserved.

(c) If $\Box\boldsymbol{\gamma} = 0$, then the corresponding Γ-sources coincide:

$$\boldsymbol{G}_2 = \boldsymbol{\Gamma}_2 = (\boldsymbol{\Gamma}_1 + \boldsymbol{\gamma}) = \boldsymbol{\Gamma}_1 + \boldsymbol{\gamma} = \boldsymbol{\Gamma}_1 = \boldsymbol{G}_1.$$

By Eqs. (16) and (17), equal Γ-sources imply equal charge and current densities: $\rho_2 = \rho_1, \boldsymbol{J}_2 = \boldsymbol{J}_1$. However, unless $\nabla\times(\nabla\times\boldsymbol{\gamma}) = 0$, the difference in Γ-potentials, $\boldsymbol{\gamma}$, leads to differences in the fields $\boldsymbol{E}$ and $\boldsymbol{B}$, as seen from Eqs. (14a) and (15).

Q.E.D.

As a representation of electromagnetic configurations, the Γ-potential inherits and elucidates the intrinsic symmetry structure embedded in classical electrodynamics, as formalized in Theorem 3. These invariances arise from the layered derivative relationships among potentials, fields, and sources. The Γ-representation provides a unified framework in which such structural freedoms are not only preserved but also rendered more transparent and systematically classified:

- When two Γ-potentials differ by a static and divergence-free vector field, they represent the same electromagnetic configuration (Theorem 3(a)).

- When the difference between the two Γ-potentials, $\boldsymbol{\gamma}$, is a solution to the homogeneous wave equation and has vanishing double curl, the scalar and vector potentials are related by a Lorenz-gauge-preserving gauge transformation, and the corresponding fields and sources remain identical (provided $\boldsymbol{B}_2 - \boldsymbol{B}_1$ decays at infinity; otherwise $\boldsymbol{B}$ may differ by a time-independent harmonic field; see Theorem 3(b)).
- When the difference $\boldsymbol{\gamma}$ satisfies the homogeneous wave equation but possesses a nonzero double curl, the Γ-sources remain unchanged, implying that charge and current densities remain the same, while the fields $\boldsymbol{E}$ and $\boldsymbol{B}$ may differ (Theorem 3(c)).
- Extending beyond Theorem 3, if the difference $\boldsymbol{\gamma}$ does not satisfy the homogeneous wave equation—i.e., if it corresponds to a nonzero Γ-source $\Box\boldsymbol{\gamma} = \boldsymbol{g} \neq 0$—then the two Γ-potentials describe genuinely distinct electromagnetic configurations, with differences in all associated quantities including charge density $\rho$ and current density $\boldsymbol{J}$.

In this way, while providing a unified description of potentials, fields, and sources, the Γ-representation also helps to reveal the layered symmetries of electromagnetic theory.

## 4. Γ-Transformations

With the mapping between EM configurations and Γ-potentials established by Theorems 1 and 2, any operation on a Γ-potential that preserves the required regularity will generate a new Γ-potential, which then maps to a new EM configuration. In this sense, transforming a Γ-potential as a vector field in principle induces a transformation of the corresponding EM configuration. This natural feature of the Γ-representation framework essentially provides a constructive tool to transform known EM solutions into new ones.

As is well known, the governing relations of classical electrodynamics—including Maxwell's equations, the charge continuity equation, the Lorenz gauge condition, the potential-field relations, and the associated wave equations (i.e., Eqs. (1)–(10))—are all linear. This linearity naturally extends to the Γ-potential framework, where the relations linking electromagnetic quantities to the Γ-potential (Eqs. (11)–(17)) are likewise linear. Consequently, applying a linear operator to a Γ-potential (e.g., scalar multiplication or superposition with another Γ-potential addition) carries these linear relationships through to the resulting Γ-potential and the corresponding EM configuration. This "linear case" transformation is thus of significant importance and forms the basis of the discussion in this section.

For convenience, we refer to a transformation of an EM configuration induced by a linear operation on the corresponding Γ-potential as a **linear Γ-transformation** (or simply a **Γ-transformation**). Transformations induced by general (linear or non-linear) operations on a Γ-potential are termed **general Γ-transformations**.

Unlike the traditional gauge transformations, which are based on internal symmetries at the potential level and alter potentials while preserving fields and sources, the Γ-transformation is based on the broader symmetries described in Section 3. It is a solution-generating map that generally modifies the entire EM configuration —including potentials, fields, and sources—while ensuring the Lorenz gauge condition, charge continuity and Maxwell's equations hold for the resulting configuration. In this context, the Γ-transformation can be considered as an extension beyond conventional gauge transformation, recovering gauge transformations as a specific, restricted subset.

### 4.1 Algebraic Operations on EM Configurations

To proceed with a more concrete discussion, we first define basic operations on electromagnetic configurations, treating them formally as algebraic objects.

Specifically, we define

- The sum of two EM solutions $[\mathrm{EM}_1] = [\varphi_1, \boldsymbol{A}_1, \boldsymbol{E}_1, \boldsymbol{B}_1, \rho_1, \boldsymbol{J}_1]$ and $[\mathrm{EM}_2] = [\varphi_2, \boldsymbol{A}_2, \boldsymbol{E}_2, \boldsymbol{B}_2, \rho_2, \boldsymbol{J}_2]$ as:

$$[\mathrm{EM}_1] + [\mathrm{EM}_2] \triangleq [\varphi_1 + \varphi_2, \boldsymbol{A}_1 + \boldsymbol{A}_2, \boldsymbol{E}_1 + \boldsymbol{E}_2, \boldsymbol{B}_1 + \boldsymbol{B}_2, \rho_1 + \rho_2, \boldsymbol{J}_1 + \boldsymbol{J}_2],$$

- The scalar multiplication of $\alpha \in \mathbb{R}$ and an EM solution $[\mathrm{EM}] = [\varphi, \boldsymbol{A}, \boldsymbol{E}, \boldsymbol{B}, \rho, \boldsymbol{J}]$ as:

$$\alpha[\mathrm{EM}] \triangleq [\alpha\varphi, \alpha\boldsymbol{A}, \alpha\boldsymbol{E}, \alpha\boldsymbol{B}, \alpha\rho, \alpha\boldsymbol{J}].$$

Under these operations, any linear combination of EM solutions remains a valid EM solution due to linearity of the governing equations. Moreover, such algebraic combinations correspond directly to linear combinations of their associated Γ-potentials, as formalized in the following theorem.

**Theorem 4** (**Γ-Potential Algebraic Linearity Theorem**).

Let $\boldsymbol{\Gamma}_1$ and $\boldsymbol{\Gamma}_2$ be two Γ-potentials corresponding to EM solutions $[\mathrm{EM}(\boldsymbol{\Gamma}_1)]$ and $[\mathrm{EM}(\boldsymbol{\Gamma}_2)]$, and let $\alpha, \beta \in \mathbb{R}$. Then:

$$[\mathrm{EM}(\alpha\boldsymbol{\Gamma}_1 + \beta\boldsymbol{\Gamma}_2)] = \alpha[\mathrm{EM}(\boldsymbol{\Gamma}_1)] + \beta[\mathrm{EM}(\boldsymbol{\Gamma}_2)].$$

**Proof.**

As Γ-potentials, $\boldsymbol{\Gamma}_1$ and $\boldsymbol{\Gamma}_2$ are both $C_t^3 C_x^3$ regular; their linear combination $\alpha\boldsymbol{\Gamma}_1 + \beta\boldsymbol{\Gamma}_2$ also preserves this regularity and thus qualifies as a Γ-potential. The remainder of the proof follows directly from the linearity of the relations linking the Γ-potential to the EM quantities (Eqs. (11)–(17)) and is omitted here.

### 4.2 Additive Γ-transformation

Theorem 4 enables the construction of new EM solutions by linearly combining existing ones through their associated Γ-potentials. Consequently, for a given EM solution $[\mathrm{EM}_0(\boldsymbol{\Gamma}_0)]$, we can generate a new solution by superimposing an auxiliary Γ-potential $\boldsymbol{\Gamma}$ onto the original $\boldsymbol{\Gamma}_0$. This yields:

$$[\mathrm{EM}'] = [\mathrm{EM}_0(\boldsymbol{\Gamma}_0)] + [\mathrm{EM}(\boldsymbol{\Gamma})] = [\mathrm{EM}(\boldsymbol{\Gamma}_0 + \boldsymbol{\Gamma})].$$

Here, $[EM']$ is formally the sum of $[\mathrm{EM}_0]$ and $[EM(\boldsymbol{\Gamma})]$. In this context, we refer to $[\mathrm{EM}']$ as the **additive Γ-transformation** of $[\mathrm{EM}_0]$ by $\boldsymbol{\Gamma}$. This is a fundamental subclass of linear Γ-transformation and is of particular interest due to its conceptual simplicity and direct applicability in field engineering.

According to Theorem 1, a spacetime vector field qualifies as a Γ-potential if it is $C_t^3 C_x^3$ regular. Any such vector field can thus be used to additively transform an EM solution. This enables a direct and flexible method for modifying specific aspects of an EM solution. In principle, we can selectively adjust a particular property of an EM solution by transforming it with a Γ-potential that induces a targeted modification of that property.

As a simple illustration, assuming we wish to cancel the magnetic field of a configuration $[EM(\boldsymbol{\Gamma}_1)]$, we can choose a vector field $\boldsymbol{\Gamma}_2$ such that $\boldsymbol{\Gamma}_1 + \boldsymbol{\Gamma}_2$ is curl-free and time-invariant. Using $\boldsymbol{\Gamma}_2$ to transform $EM(\boldsymbol{\Gamma}_1)$, the resulting solution $[\mathrm{EM}'] = [\mathrm{EM}(\boldsymbol{\Gamma}_1)] + [\mathrm{EM}(\boldsymbol{\Gamma}_2)] = [EM(\boldsymbol{\Gamma}_1 + \boldsymbol{\Gamma}_2)]$ then has a vanishing magnetic field (as per Eq. (15)), while other components are systematically reconfigured.

Furthermore, any sufficiently regular vector function of the quantities within a given EM solution can serve as a Γ-potential for transformation. This opens the door to recursive or self-referential transformations. For example, taking the electric field $\boldsymbol{E}$ from a solution $[EM_0\ ]$ and using its vector function $\boldsymbol{F}(\boldsymbol{E})$ to transform the same solution yields:

$$[EM'] = [EM_0\ ] + \left[EM\big(\boldsymbol{F}(\boldsymbol{E})\big)\right]$$

Here, the new configuration $[EM']$ may thus incorporate feedback-like effects, potentially enhancing or reorienting the original electric field in $[EM_0\ ]$. If the initial field $\boldsymbol{E}$ is time-dependent and spatially localized, the transformed field may exhibit intensified radiation characteristics or constructive interference in specific directions. Similarly, using vector functions of the magnetic field $\boldsymbol{B}$ or the vector potential $\boldsymbol{A}$ to transform an EM solution may reveal or enhance the topological features such as helicity. While a full exploration of such self-referential maps is beyond the scope of this paper, they provide an avenue for future research.

In passing, these transformations can also be combined or iterated systematically. For instance, applying successive transformations with a fixed Γ-potential,

$$\left[\mathrm{EM}^{(\mathrm{n})}\right] = \left[\mathrm{EM}^{(\mathrm{n}-1)}\right] + [\mathrm{EM}(\boldsymbol{\Gamma})]$$

generates a sequence of solutions $\left[\mathrm{EM}^{(0)}\right]$, $\left[\mathrm{EM}^{(1)}\right]$,…, $\left[\mathrm{EM}^{(\mathrm{n})}\right]$. The cumulative effect may include the progressive buildup of energy, helicity, or spatial complexity, providing a pathway for studying the synthesis and evolution of complex electromagnetic configurations.

### 4.3 Linear Operator-Induced Γ-transformations

Beyond the algebraic operations discussed in Sections 4.1 and 4.2, the Γ-representation framework also supports transformations induced by general linear operators (such as differential or integral operators) acting on the Γ-potential. This provides an extended mechanism for generating new electromagnetic configurations with further modified structural characteristics.

Owing to the linearity of the defining relations and of the wave equation, general linear operators (beyond the basic algebraic operations) can also carry structured relationships from Γ-potentials through to the resulting EM quantities. This principle is formalized in the following theorem:

**Theorem 5 (Linear Operator-Induced Γ-Transformation Theorem).**

Let $\boldsymbol{\Gamma}(\boldsymbol{x},t) \in C_t^3 C_x^3(\mathbb{R}\times\mathbb{R}^3;\mathbb{R}^3)$ be a Γ-potential associated with $[\mathrm{EM}(\boldsymbol{\Gamma})] = [\varphi,\boldsymbol{A},\boldsymbol{E},\boldsymbol{B},\rho,\boldsymbol{J}]$, and $\Box\boldsymbol{\Gamma} = \boldsymbol{G}$. Let $\mathcal{L}$ be a linear operator on vector fields such that, for this specific $\boldsymbol{\Gamma}$, $\mathcal{L}[\boldsymbol{\Gamma}] \in C_t^3 C_x^3$. Assume that $\mathcal{L}$ commutes with temporal and spatial derivatives as well as the wave operator:

$$[\mathcal{L},\partial_t] = [\mathcal{L},\nabla\cdot] = [\mathcal{L},\nabla\times] = \left[\mathcal{L},\Box\right] = 0, \text{ (where } [A,B] = AB - BA\text{)}.$$

Furthermore, assume there exists a linear scalar companion operator $\mathcal{L}_s$ such that, for any sufficiently smooth vector field $\boldsymbol{X}$ and scalar function $\psi$,

$$\nabla\cdot(\mathcal{L}[\boldsymbol{X}]) = \mathcal{L}_s[\nabla\cdot\boldsymbol{X}], \quad \mathcal{L}[\nabla\psi] = \nabla(\mathcal{L}_s[\psi]).$$

Then, the EM configuration associated with $\mathcal{L}[\boldsymbol{\Gamma}]$, denoted by $[\mathrm{EM}(\mathcal{L}[\boldsymbol{\Gamma}])] = [\varphi', \boldsymbol{A}', \boldsymbol{E}', \boldsymbol{B}', \rho', \boldsymbol{J}']$, is given by:

$$\varphi' = \mathcal{L}_s[\varphi],\ \boldsymbol{A}' = \mathcal{L}[\boldsymbol{A}],\ \boldsymbol{E}' = \mathcal{L}[\boldsymbol{E}],\ \boldsymbol{B}' = \mathcal{L}[\boldsymbol{B}],\ \rho' = \mathcal{L}_s[\rho],\ \boldsymbol{J}' = \mathcal{L}[\boldsymbol{J}],$$

and the transformed Γ-source is $\boldsymbol{G}' = \mathcal{L}[\boldsymbol{G}]$.

The proof follows directly from the linearity of the defining relations and the stated commutation and compatibility properties; it is omitted here for brevity.

Depending on the particular operator $\mathcal{L}$, $\boldsymbol{\Gamma}$ may need higher regularity to ensure $\mathcal{L}[\boldsymbol{\Gamma}] \in C_t^3 C_x^3$ (e.g., for a spacetime operator of order $(m_t, m_x)$, it suffices that $\boldsymbol{\Gamma} \in C_t^{3+m_t} C_x^{3+m_x}$ ). Under the structural assumptions on $\mathcal{L}$ and $\mathcal{L}_s$, the scalar outputs $\mathcal{L}_s[\varphi]$ and $\mathcal{L}_s[\rho]$ inherit the required scalar regularity.

As a simple illustration of such transformations, if for a Γ-potential $\boldsymbol{\Gamma} \in C_t^3 C_x^4$ we take the curl operator $\mathcal{L} = \nabla \times$, then $\boldsymbol{\Gamma}' = \mathcal{L}[\boldsymbol{\Gamma}] = \nabla \times \boldsymbol{\Gamma}$, $\mathcal{L}_s \equiv 0$, and the transformed solution $[\mathrm{EM}(\mathcal{L}[\boldsymbol{\Gamma}])]$ satisfies:

$$\varphi' = 0,\ \boldsymbol{A}' = \nabla \times \boldsymbol{A} = \boldsymbol{B},\ \boldsymbol{E}' = \nabla \times \boldsymbol{E},\ \boldsymbol{B}' = \nabla \times \boldsymbol{B} = \nabla \times (\nabla \times \boldsymbol{A}),\ \rho' = 0,\ \boldsymbol{J}' = \nabla \times \boldsymbol{J},\ \boldsymbol{G}' = \nabla \times \boldsymbol{G}.$$

We can see here that taking the curl of the Γ-potential induces, in effect, the curl of the entire electromagnetic configuration. In other words, the curl operation applied to $\boldsymbol{\Gamma}$ is carried through to all EM quantities—potentials, fields, and sources. Consequently, the components not tied to curl (i.e., the longitudinal/gradient part) are effectively filtered out in the transformed configuration—e.g., $\varphi$ and $\rho$ vanish—leaving only the solenoidal (divergence-free) content.

Particularly, this operator transforms the original magnetic field B into the vector potential $\boldsymbol{A}'$ in the new configuration. This follows directly from the classical relation $\boldsymbol{B} = \nabla \times \boldsymbol{A}$: under $\mathcal{L} = \nabla \times$, we obtain $\boldsymbol{A}' = \nabla \times \boldsymbol{A} = \boldsymbol{B}$. In this sense, such an operator "shifts" the representation to a higher-order curl level, promoting curl-generated field content into the role of potentials while filtering out purely longitudinal components. Furthermore, repeating this transformation generates a hierarchy of configurations built from successive curl operations, revealing a ladder structure within the Γ-representation.

For brevity, we may write $[\mathrm{EM}'] = \nabla \times [\mathrm{EM}]$, or more generally, $[\mathrm{EM}'] = \mathcal{L}[\mathrm{EM}]$. Equivalently, Theorem 5 can be expressed as $[\mathrm{EM}(\mathcal{L}[\boldsymbol{\Gamma}])] = \mathcal{L}[\mathrm{EM}(\boldsymbol{\Gamma})]$.

Under the hypotheses of Theorem 5, finite compositions of admissible operators are also admissible (subject to the regularity condition). Examples include the vector Laplacian $\nabla^2$, the d'Alembert operator , the time derivative $\partial_t$, and isotropic scalar filters $f(\nabla^2)$ or $f(\partial_t, \nabla^2)I$ (applied component-wise), and mixed forms such as $f(\partial_t, \nabla^2)I + g(\partial_t, \nabla^2)(\nabla \times)$.

These higher-order, structure-preserving transformations can be useful when vortex-like structures or other topological features are of interest.

## 5. Algebraic Perspective on Structure of the Electromagnetic Solution Space

Building on the linear structure established in Section 4, we now explore the algebraic structures of the electromagnetic solution space as revealed through the Γ-representation.

From Theorems 1 and 2, every regular vacuum EM solution corresponds uniquely to a Γ-potential (modulo the symmetries described in Theorem 3), and vice versa. Thus, the space of all EM solutions—denoted $\mathcal{E}$—can be viewed as isomorphic to the vector space

$$\mathcal{V} \triangleq C_t^3 C_x^3(\mathbb{R} \times \mathbb{R}^3; \mathbb{R}^3)$$

modulo an equivalence relation $\sim$ induced by symmetries:

$$\mathcal{E} \cong \mathcal{V}/\sim.$$

The equivalence relation $\sim$ depends on which physical attributes of the EM solution are deemed invariant under transformation. Theorem 3 identifies a hierarchy of such symmetries, each inducing a distinct quotient structure on the vector field space $\mathcal{V}$:

- **Identity Equivalence:** If two Γ-potentials differ by a static, divergence-free field, then they yield exactly the same potentials, fields, and sources. This defines the finest equivalence relation (Theorem 3(a)), and the resulting quotient $(\mathcal{V}/\sim)_a$ classifies Γ-potentials that represent identical EM solutions in full.
- **Gauge Equivalence:** If we regard two EM solutions as equivalent whenever they produce the same fields and sources—even if their scalar and vector potentials differ via a gauge transformation—then the relevant equivalence relation (Theorem 3(b)) is coarser. The quotient space $(\mathcal{V}/\sim)_b$ corresponds to field-source equivalence under gauge symmetry (preserving the Lorenz gauge; with standard decay at infinity for $\boldsymbol{B}_2 - \boldsymbol{B}_1$).
- **Source Equivalence:** Finally, if we consider two solutions equivalent whenever they originate from the same source $(\rho, \boldsymbol{J})$, we obtain the coarsest equivalence class (Theorem 3(c)). The quotient $(\mathcal{V}/\sim)_c$ captures source-preserving transformations, under which the fields may differ due to the allowed addition of source-free solutions (i.e., free-space wavefields).

In a broader algebraic context, these nested equivalence relations suggest that the solution space can be analyzed as a representation of various symmetry groups acting on $\mathcal{V}$. Under this view, the quotient structures $(\mathcal{V}/\sim)$ correspond to the orbits of the solution space under the action of specific operator groups. This group-theoretic framing may provide a unified language for studying how electromagnetic properties remain invariant or evolve under different classes of transformations.

The isomorphic relation between the EM solution space and the vector field space $\mathcal{V}$ (modulo symmetry) also invites the application of well-established results from vector field analysis to the study of electromagnetic solutions.

As an illustration, the Helmholtz decomposition asserts that any $\boldsymbol{\Gamma} \in \mathcal{V}$ can be expressed as the sum of a curl-free part and a divergence-free part (under suitable boundary conditions or sufficient decay at infinity). Applied to a Γ-potential, we write:

$$\boldsymbol{\Gamma} = -\nabla\psi + \nabla \times \boldsymbol{C} \triangleq \boldsymbol{\Gamma}_{irr} + \boldsymbol{\Gamma}_{sol}.$$

This leads to a decomposition of the corresponding EM configuration:

$$[\mathrm{EM}(\boldsymbol{\Gamma})] = [\mathrm{EM}(\boldsymbol{\Gamma}_{irr})] + [\mathrm{EM}(\boldsymbol{\Gamma}_{sol})]$$

By Eqs. (12)–(17), $[\mathrm{EM}(\boldsymbol{\Gamma}_{irr})]$ (the curl-free component) is purely electric, with $\boldsymbol{B}_{\Gamma_{irr}} = 0$; whereas $[\mathrm{EM}(\boldsymbol{\Gamma}_{sol})]$ (the divergence-free component) satisfies $\varphi_{\Gamma_{sol}} = 0$ and $\rho_{\Gamma_{sol}} = 0$ (Eqs. (11, 16)). In this way, an EM solution is decomposed into two physically distinct components: one capturing the longitudinal or potential-driven aspects, and the other the transverse or rotational aspects. Such a decomposition may facilitate the analysis of the physical origin of fields while offering a flexible means to isolate or synthesize particular features in complex electromagnetic environments.

Beyond decomposition, the same framework may support the construction of topologically structured solutions—for example, those with knotted or linked field lines—by engineering Γ-potentials to inherit prescribed topological invariants from known vector field configurations.

In this sense, the Γ-representation endows the set of electromagnetic solutions with a natural vector-space structure. The layered quotients precisely encode which physical distinctions are retained in a given context, linking the analysis of Maxwell's equations to the algebra of vector fields.

## 6. Revisiting the Polarized Media Case—with Connections between Γ-Representation and the Hertz Potentials

As an application of the Γ-representation, we revisit the structure of electromagnetic fields in media—a classical setting in which the Hertz potentials are frequently applied. This offers a brief comparison between the Γ-potential and the classical Hertz potentials, illustrating where they coincide and where they diverge.

### 6.1 Effective Source Pairs in Polarized Media

In isotropic, homogeneous, linear media, the constitutive relations are given by:

$$\boldsymbol{D} = \varepsilon \boldsymbol{E} = \varepsilon_0 \boldsymbol{E} + \boldsymbol{P}, \quad \boldsymbol{H} = \frac{1}{\mu}\boldsymbol{B} = \frac{1}{\mu_0}\boldsymbol{B} - \boldsymbol{M}, \tag{26}$$

where $\boldsymbol{P}$ and $\boldsymbol{M}$ are the electric polarization and magnetization vectors. In a linear medium, they are parallel and proportional to $\boldsymbol{E}$ and $\boldsymbol{B}$ respectively. From Eq. (26), we have

$$\boldsymbol{P} = (\varepsilon - \varepsilon_0)\boldsymbol{E}, \quad \boldsymbol{M} = \left(\frac{1}{\mu_0} - \frac{1}{\mu}\right)\boldsymbol{B}. \tag{27}$$

As Stratton shows (*Electromagnetic Theory* [3], Sec. 1.6), Maxwell's equations in such media can be written as

$$\begin{gathered} \nabla \cdot \boldsymbol{E} = \frac{1}{\varepsilon_0}(\rho_0 - \nabla \cdot \boldsymbol{P}), \quad \nabla \cdot \boldsymbol{B} = 0, \\ \nabla \times \boldsymbol{E} = -\frac{\partial \boldsymbol{B}}{\partial t}, \quad \nabla \times \boldsymbol{B} = \frac{1}{c^2}\frac{\partial \boldsymbol{E}}{\partial t} + \mu_0\left(\boldsymbol{J}_0 + \frac{\partial \boldsymbol{P}}{\partial t} + \nabla \times \boldsymbol{M}\right). \end{gathered} \tag{28}$$

In alignment with this formulation, Stratton states [3]: *"the presence of rigid material bodies in an electromagnetic field may be completely accounted for by an equivalent distribution of charge of density* $-\nabla \cdot \boldsymbol{P}$*, and an equivalent distribution of current of density* $\frac{\partial \boldsymbol{P}}{\partial t} + \nabla \times \boldsymbol{M}$*."*

Accordingly, electromagnetic fields in such media are equivalent to fields (in vacuum) with the following total effective charge and current densities, which satisfy the continuity equation (Eq. (1)) and thus constitute a valid source pair:

$$\rho = \rho_0 - \nabla \cdot \boldsymbol{P}\,, \quad \boldsymbol{J} = \boldsymbol{J}_0 + \frac{\partial \boldsymbol{P}}{\partial t} + \nabla \times \boldsymbol{M} \tag{29}$$

Leveraging the linearity of EM configurations, we can decompose this effective source pair $(\rho, \boldsymbol{J})$ into three separate groups:

- Electric polarization sources: $\left(\rho_e = -\nabla \cdot \boldsymbol{P}, \boldsymbol{J}_e = \frac{\partial \boldsymbol{P}}{\partial t}\right)$;
- Magnetization sources: $(\rho_m = 0, \boldsymbol{J}_m = \nabla \times \boldsymbol{M})$;
- Free sources: $(\rho_0, \boldsymbol{J}_0)$.

Each group satisfies the continuity equation individually, and hence defines a valid source pair. If we solve the fields with each pair separately and superpose the results, we obtain the total solution of fields in the media. This is a common route in the classical Hertz-potential treatment of polarized media. Within this route, the Γ-representation provides a parallel, equivalent scheme that leads to exactly the same physical results while offering an alternative perspective on the classical Hertz potentials. Specifically, the Γ-potential can be shown to match the electric Hertz potential (up to convention-dependent constants), while the magnetic Hertz potential can be obtained from a Γ-potential through a specific operator-induced transformation described by Theorem 5.

Furthermore, the Γ-representation allows for an alternative, more direct and uniform approach: starting from the total effective sources $(\rho, J)$ in Eq. (29), construct a single Γ-source $\mathbf{G}$ from Eqs. (16)–(17) and solve the vector wave equation to obtain the corresponding Γ-potential, from which the full EM solution follows via the Γ-representation formulas.

The remainder of this section is devoted to discussing these different approaches in detail.

### 6.2 Electric Polarization Sources: Similarity between the Γ-potential and the Electric Hertz potential

The electric Hertz potential, $\boldsymbol{\Pi}_e$, is essentially dedicated to the treatment of electric polarization. However, depending on the theoretical perspective adopted, its definition can vary slightly in the literature.

When electromagnetic radiation is the primary focus (e.g., in Sec. 20.2 of [7]), $\boldsymbol{\Pi}_e$ is defined under a "vacuum-perspective" convention, utilizing the vacuum constants $\varepsilon_0$ and $\mu_0$ (and consequently, light speed in vacuum $c = 1/\sqrt{\varepsilon_0 \mu_0}$ ):

$$\varphi = -\nabla \cdot \boldsymbol{\Pi}_e, \quad \boldsymbol{A} = \frac{1}{c^2}\frac{\partial \boldsymbol{\Pi}_e}{\partial t}, \quad \frac{1}{c^2}\frac{\partial^2 \boldsymbol{\Pi}_e}{\partial t^2} - \nabla^2 \boldsymbol{\Pi}_e = \frac{\boldsymbol{P}}{\varepsilon_0}, \tag{30}$$

and the related electromagnetic fields are likewise formulated in terms of these vacuum constants:

$$\boldsymbol{E}_e = -\varepsilon_0 \mu_0 \frac{\partial^2 \boldsymbol{\Pi}_e}{\partial t^2} + \nabla(\nabla \cdot \boldsymbol{\Pi}_e) = \nabla \times (\nabla \times \boldsymbol{\Pi}_e) - \frac{\boldsymbol{P}}{\varepsilon_0}, \quad \boldsymbol{B}_e = \frac{1}{c^2}\frac{\partial}{\partial t}(\nabla \times \boldsymbol{\Pi}_e). \tag{31}$$

Conversely, when the analysis centers on fields within a material medium (e.g., in [5], [6]), a "medium-perspective" convention is employed, defining $\boldsymbol{\Pi}_e$ with the medium parameters $\varepsilon$ and $\mu$ instead of the vacuum ones:

$$\varphi = -\nabla \cdot \boldsymbol{\Pi}_e, \quad \boldsymbol{A} = \varepsilon\mu \frac{\partial \boldsymbol{\Pi}_e}{\partial t}, \quad \varepsilon\mu \frac{\partial^2 \boldsymbol{\Pi}_e}{\partial t^2} - \nabla^2 \boldsymbol{\Pi}_e = \frac{\boldsymbol{P}}{\varepsilon}. \tag{32}$$

and the related fields are:

$$\boldsymbol{E}_e = -\varepsilon\mu \frac{\partial^2 \boldsymbol{\Pi}_e}{\partial t^2} + \nabla(\nabla \cdot \boldsymbol{\Pi}_e) = \nabla \times (\nabla \times \boldsymbol{\Pi}_e) - \frac{\boldsymbol{P}}{\varepsilon}, \quad \boldsymbol{B}_e = \varepsilon\mu \frac{\partial}{\partial t}(\nabla \times \boldsymbol{\Pi}_e). \tag{33}$$

Mathematically, the electric Hertz potential $\boldsymbol{\Pi}_e$ defined by Eq. (32) is, in general, different from that defined by Eq. (30), as the two definitions employ different wave operators. Physically, however, they are equivalent: for the same polarization distribution and compatible boundary/radiation conditions, the resulting electromagnetic fields $\boldsymbol{E}$ and $\boldsymbol{B}$ obtained from either definition (i.e., Eq. (31) or Eq. (33)) are identical.

Turning now to the Γ-representation formalism, for the electric polarization source pair $(\rho_e, \boldsymbol{J}_e) = \left(-\nabla \cdot \boldsymbol{P}, \frac{\partial \boldsymbol{P}}{\partial t}\right)$, if we define a vector $\boldsymbol{G}_e = \boldsymbol{P}/\varepsilon_0$, it naturally satisfies the consistency equations (Eqs. (16) and (17)) together with $\rho_e$ and $\boldsymbol{J}_e$. Thus, $\boldsymbol{G}_e$ qualifies as the Γ-source for this source pair, and the corresponding vector wave equation Eq. (11) becomes:

$$\varepsilon_0\mu_0 \frac{\partial^2 \boldsymbol{\Gamma}_e}{\partial t^2} - \nabla^2 \boldsymbol{\Gamma}_e = \boldsymbol{G}_e = \frac{\boldsymbol{P}}{\varepsilon_0} \tag{34}$$

Solving this for the Γ-potential $\boldsymbol{\Gamma}_e$, the resulting scalar and vector potentials and the associated fields are given by Eqs. (12)-(15) as:

$$\varphi_e = -\nabla \cdot \boldsymbol{\Gamma}_e, \quad \boldsymbol{A}_e = \frac{1}{c^2} \frac{\partial \boldsymbol{\Gamma}_e}{\partial t}, \tag{35}$$

$$\boldsymbol{E}_e = -\varepsilon_0\mu_0 \frac{\partial^2 \boldsymbol{\Gamma}_e}{\partial t^2} + \nabla(\nabla \cdot \boldsymbol{\Gamma}_e) = \nabla \times (\nabla \times \boldsymbol{\Gamma}_e) - \frac{\boldsymbol{P}}{\varepsilon_0}, \boldsymbol{B}_e = \frac{1}{c^2} \frac{\partial}{\partial t}(\nabla \times \boldsymbol{\Gamma}_e). \tag{36}$$

Comparing term by term, Eqs. (34)-(36) are precisely equivalent to Eqs. (30), (31). The Γ-potential in this configuration is formally identical to the electric Hertz potential defined from the vacuum-perspective. In this context, the two formalisms merge into a single mathematical description of the polarization-induced fields.

Extensions of the *electric* Hertz-potential formalism to include free sources have been discussed in the literature. Along this route, Group 3 (the free-source pair $(\rho_0, \boldsymbol{J}_0)$) can be treated within the same electric-Hertz framework by adopting its source-inclusive form, in which the polarization source pair $(\rho_e, \boldsymbol{J}_e)$ is replaced by $(\rho_0, \boldsymbol{J}_0)$. In parallel, the Γ-framework is source-inclusive from the outset and offers an equivalent route. The two formalisms yield the same physical solution for the same free-source contribution.

In this sense, the Γ-representation provides a parallel description that aligns with the classical electric Hertz potential framework for polarization-induced and free source configurations.

It is worth noting that, due to the source equivalence described in Section 5, the fields associated with any specific source pair are unique only up to a source-free contribution. This principle holds for all three source groups, regardless of the methodology employed (i.e., the Hertz potential or the Γ-representation formalism). The total homogeneous contributions from the three groups are ultimately determined by the boundary and initial conditions.

### 6.3 Magnetization in Media: Transformation between Γ-potential and the Magnetic Hertz potential

For Group 2 (magnetization sources), the effective source pair is $(\rho_m, \boldsymbol{J}_m) = (0,\ \nabla \times \boldsymbol{M})$. Recall from Eq. (27) that $\boldsymbol{M}$ is proportional to $\boldsymbol{B}$ with a constant coefficient. $\boldsymbol{M}$ hence inherits the regularity of $\boldsymbol{B}$. Furthermore, from $\nabla \cdot \boldsymbol{B} = 0$ (Eq. (3)), we have $\nabla \cdot \boldsymbol{M} = 0$.

Given the charge and current densities, we can proceed as in the electric polarization case: construct a Γ-source from $\rho_m$ and $\boldsymbol{J}_m$, solve the wavefield equation to determine the Γ-potential, then subsequently derive the complete electromagnetic solution.

Specifically, if we define a vector field

$$\boldsymbol{G}_m = \frac{1}{\varepsilon_0}\int_{t_0}^{t} (\nabla \times \boldsymbol{M}) d\tau, \tag{37}$$

and adopt it as a Γ-source, the consistency equations (Eqs. (16)–(17)) are satisfied for $(\rho_m,\ \boldsymbol{J}_m)$:

$$\varepsilon_0 \frac{\partial \boldsymbol{G}_m}{\partial t} = \nabla \times \boldsymbol{M} = \boldsymbol{J}_m, \quad \nabla \cdot \boldsymbol{G}_m = 0 = \rho_m.$$

The governing wave equation for the corresponding Γ-potential $\boldsymbol{\Gamma}_m$ is then:

$$\frac{1}{c^2}\frac{\partial^2 \boldsymbol{\Gamma}_m}{\partial t^2} - \nabla^2 \boldsymbol{\Gamma}_m = \boldsymbol{G}_m = \frac{1}{\varepsilon_0}\int_{t_0}^{t} (\nabla \times \boldsymbol{M}) d\tau. \tag{38}$$

Solving (38) for $\boldsymbol{\Gamma}_m$, the full solution then follows from Eqs. (12)–(17).

Within the Γ-representation formalism, an alternative treatment exists. This route is equivalent to the previous one via an operator-induced transformation and runs in exact parallel with the magnetic Hertz potential approach.

Consider an auxiliary Γ-source $\boldsymbol{G}_a = \mu_0 \boldsymbol{M}$, and let the corresponding (auxiliary) Γ-potential (denoted as $\boldsymbol{\Gamma}_a$) solve:

$$\frac{1}{c^2}\frac{\partial^2 \boldsymbol{\Gamma}_a}{\partial t^2} - \nabla^2 \boldsymbol{\Gamma}_a = \boldsymbol{G}_a = \mu_0 \boldsymbol{M}. \tag{39}$$

The associated (auxiliary) EM solution $[\mathrm{EM}_a] = [\varphi_a, \boldsymbol{A}_a, \boldsymbol{E}_a, \boldsymbol{B}_a, \rho_a, \boldsymbol{J}_a]$ is then given by Eqs. (12)–(17) as:

$$\varphi_a = -\nabla \cdot \boldsymbol{\Gamma}_a, \boldsymbol{A}_a = \frac{1}{c^2}\frac{\partial \boldsymbol{\Gamma}_a}{\partial t}, \boldsymbol{E}_a = -\frac{1}{c^2}\frac{\partial^2 \boldsymbol{\Gamma}_a}{\partial t^2} + \nabla(\nabla \cdot \boldsymbol{\Gamma}_a)$$

$$\boldsymbol{B}_a = \frac{1}{c^2}\frac{\partial}{\partial t}(\nabla \times \boldsymbol{\Gamma}_a), \rho_a = 0, \boldsymbol{J}_a = \frac{1}{c^2}\frac{\partial \boldsymbol{M}}{\partial t}. \tag{40}$$

Now, define a linear operator

$$\mathcal{L} \triangleq c^2 \left( \int_{t_0}^{t} \cdot \, d\tau \right) \circ (\nabla \times) \tag{41}$$

Assuming $t_0 = -\infty$ with causal decay (or impose zero initial data at a finite $t_0$), $\mathcal{L}$ is compatible with the differential operators in the sense required by Theorem 5: it commutes with spatial derivatives (e.g., $\nabla \times$), and with $\partial_t$ (and hence with $\Box$) on the class of fields considered here. Moreover, $\mathcal{L}$ admits the scalar companion $\mathcal{L}_s \equiv 0$ (since $\nabla \cdot (\mathcal{L}[\boldsymbol{X}]) = 0$ and $\mathcal{L}[\nabla\psi] = 0$ for sufficiently smooth $\boldsymbol{X}$ and $\psi$). Hence, $\mathcal{L}$ (together with its companion $\mathcal{L}_s$) induces a Γ-transformation as described by Theorem 5.

Remarkably, this transformation maps the auxiliary Γ-source $\boldsymbol{G}_a$ exactly to the magnetization Γ-source $\boldsymbol{G}_m$ defined in Eq. (37):

$$\mathcal{L}[\boldsymbol{G}_a] = c^2 \int_{t_0}^{t} \big(\nabla \times (\mu_0 \boldsymbol{M})\big) d\tau = \frac{1}{\varepsilon_0} \int_{t_0}^{t} (\nabla \times \boldsymbol{M}) d\tau = \boldsymbol{G}_m.$$

Consequently, the auxiliary Γ-potential $\boldsymbol{\Gamma}_a$ is transformed to $\boldsymbol{\Gamma}_m$ (i.e., $\mathcal{L}[\boldsymbol{\Gamma}_a] = \boldsymbol{\Gamma}_m$). Following Theorem 5, the entire EM solution $EM(\boldsymbol{\Gamma}_a)$ is transformed to the EM solution associated to $\boldsymbol{\Gamma}_m$, i.e.,

$$EM(\boldsymbol{\Gamma}_m) = \mathcal{L}[EM(\boldsymbol{\Gamma}_a)].$$

In other words, applying $\mathcal{L}$ and $\mathcal{L}_s$ to the potentials, fields and sources in Eq. (40) transforms them to their counterparts associated to $\boldsymbol{\Gamma}_m$, i.e.,

$$\begin{gathered}
\varphi_m = \mathcal{L}_s[\varphi_a] = 0, \\
\boldsymbol{A}_m = \mathcal{L}[\boldsymbol{A}_a] = \mathcal{L}\left[\frac{1}{c^2}\frac{\partial \boldsymbol{\Gamma}_a}{\partial t}\right] = \nabla \times \int_{t_0}^{t} \frac{\partial \boldsymbol{\Gamma}_a}{\partial t} d\tau = \nabla \times \boldsymbol{\Gamma}_a, \\
\boldsymbol{E}_m = \mathcal{L}[\boldsymbol{E}_a] = \mathcal{L}\left[-\frac{1}{c^2}\frac{\partial^2 \boldsymbol{\Gamma}_a}{\partial t^2} + \nabla(\nabla \cdot \boldsymbol{\Gamma}_a)\right] = -\frac{\partial}{\partial t}(\nabla \times \boldsymbol{\Gamma}_a), \\
\boldsymbol{B}_m = \mathcal{L}[\boldsymbol{B}_a] = \mathcal{L}\left[\frac{1}{c^2}\frac{\partial}{\partial t}(\nabla \times \boldsymbol{\Gamma}_a)\right] = \nabla \times \nabla \times \boldsymbol{\Gamma}_a, \\
\rho_m = \mathcal{L}_s[\rho_a] = 0, \\
\boldsymbol{J}_m = \mathcal{L}[\boldsymbol{J}_a] = \mathcal{L}\left[\frac{1}{c^2}\frac{\partial \boldsymbol{M}}{\partial t}\right] = \nabla \times \boldsymbol{M}.
\end{gathered}$$

Turning to the Hertz potential formalism, due to the special properties of the magnetization source pair $(\rho_m, \boldsymbol{J}_m) = (0,\ \nabla \times \boldsymbol{M})$ (a vanishing charge density and a solenoidal current), a dedicated magnetic Hertz potential $\boldsymbol{\Pi}_m$ is defined to treat this case.

As magnetization happens inside the medium, the magnetic Hertz potential $\boldsymbol{\Pi}_m$ is typically defined in the medium-perspective, i.e., using the medium parameters $\varepsilon$ and $\mu$ in most of the literature (e.g., in [5], [6]). It can, in principle, be formulated in the vacuum-perspective using the vacuum constants $\varepsilon_0$ and $\mu_0$ (e.g., in [7]). In that case, $\boldsymbol{\Pi}_m$ satisfies:

$$\varphi = 0, \boldsymbol{A} = \nabla \times \boldsymbol{\Pi}_m, \varepsilon_0 \mu_0 \frac{\partial^2 \boldsymbol{\Pi}_m}{\partial t^2} - \nabla^2 \boldsymbol{\Pi}_m = \mu_0 \boldsymbol{M}.$$

and the corresponding fields are

$$\boldsymbol{E} = -\frac{\partial}{\partial t}(\nabla \times \boldsymbol{\Pi}_m), \boldsymbol{B} = \nabla \times \nabla \times \boldsymbol{\Pi}_m.$$

Comparing these with the auxiliary Γ-potential $\boldsymbol{\Gamma}_a$, it is apparent that $\boldsymbol{\Pi}_m$ and $\boldsymbol{\Gamma}_a$ are formally identical. Both are governed by the same wave equation with $\boldsymbol{M}$ as the source and yield identical fields.

In summary, the Γ-representation offers two parallel, equivalent routes for treating magnetization: one via a direct source construction and another via an operator-induced transformation. The latter proves to be formally equivalent to the magnetic Hertz potential method. All these frameworks lead to the same physical solution associated to the magnetization source pair $(\rho_m, \boldsymbol{J}_m)$.

### 6.4 Solving with Total Effective Sources in Media: A Unified Γ-Potential Approach

Unlike in typical Hertz potential formalisms, which treat electric polarization and magnetization separately using electric and magnetic Hertz potentials, the Γ-representation enables a uniform treatment directly based on the total effective source pair. This permits a single-step approach: directly construct a total Γ-source from the ensemble of effective sources in the medium, solve for a single total Γ-potential, and then recover the entire EM solution.

Specifically, following the definition of total effective charge and current densities in Eq. (29), $\rho = \rho_0 - \nabla \cdot \boldsymbol{P}$, $\boldsymbol{J} = \boldsymbol{J}_0 + \frac{\partial \boldsymbol{P}}{\partial t} + \nabla \times \boldsymbol{M}$, we can define a total Γ-source, $\boldsymbol{G}_{tot}$, as:

$$\boldsymbol{G}_{tot} = \boldsymbol{G}_0 + \frac{\boldsymbol{P}}{\varepsilon_0} + \frac{1}{\varepsilon_0}\int_{t_0}^{t} (\nabla \times \boldsymbol{M}) d\tau$$

where $\boldsymbol{G}_0$ is a Γ-source corresponding to the free source pair $(\rho_0, \boldsymbol{J}_0)$. By construction, $\boldsymbol{G}_{tot}$ satisfies the consistency relations (Eqs. (16)-(17)) with respect to the total sources $\rho$ and $\boldsymbol{J}$. A single total Γ-potential $\boldsymbol{\Gamma}_{tot}$ is then obtained by solving the unified vector wave equation:

$$\frac{1}{c^2}\frac{\partial^2 \boldsymbol{\Gamma}_{tot}}{\partial t^2} - \nabla^2 \boldsymbol{\Gamma}_{tot} = \boldsymbol{G}_{tot}.$$

Once $\boldsymbol{\Gamma}_{tot}$ is determined, the complete electromagnetic solution in the medium is recovered through Eqs. (12)–(17). The resulting solution is identically consistent with the superposition of individual solutions, whether derived through the Γ-representation or the Hertz potential framework.

In this sense, this integrated route is distinctive to the Γ-representation: it uses a single Γ-source $\boldsymbol{G}$ as a unified vector-wavefield entry point that encodes the full effective source pair $(\rho, \boldsymbol{J})$. This helps to bypass the need to switch between distinct potential types for different source groups, while remaining fully consistent with the superposition principle. In standard Hertz-potential treatments, the corresponding construction is usually presented through separate electric and magnetic Hertz potentials combined at the level of scalar/vector potentials or the fields.

## 7. A Further Application Example: Exact Inverse-Source Construction for a Gaussian Wave Packet

The Γ-framework provides a constructive route for generating exact Maxwell-consistent configurations together with self-consistent sources. As an illustrative example, we consider an inverse-source construction for a spatiotemporally localized (rapidly decaying) wave packet in vacuum. Similar constructions can be carried out within other formulations, often through explicit evaluation of retarded convolution integrals. However, in this transient, localized setting, the retarded-integral route typically induces nonseparable spatiotemporal coupling, making the analysis analytically unwieldy in practice. The Γ-representation, on the other hand, offers a practical alternative: it proceeds by local differentiation in real spacetime (avoiding the retarded-potential integrals) while naturally respecting the Lorenz-gauge structure and charge conservation at the level of the representation.

**7.1. Exact inverse-source construction via the Γ-potential**

To construct a Gaussian wave packet (a pulsed, rapidly decaying spatiotemporal envelope, not a monochromatic Gaussian beam), we start from a prescribed Γ-potential representing a linearly polarized packet with spatial width $\sigma$ and temporal duration $\tau$:

$$\boldsymbol{\Gamma}(r,t) = \hat{\boldsymbol{z}}\Gamma_0 \exp\left(-\frac{r^2}{\sigma^2} - \frac{t^2}{\tau^2}\right), \tag{42}$$

where $\Gamma_0$ is a constant scalar amplitude.

This function is $C^\infty$ smooth and vanishes rapidly at infinity, thus satisfying the regularity requirements of Theorem 1 and thereby ensuring that all derived quantities are well-defined classical functions.

The corresponding Γ-source $\boldsymbol{G}$**,** which encodes the necessary source distribution, is determined by the wave operator:

$$\boldsymbol{G}(\boldsymbol{r},t) = \Box\boldsymbol{\Gamma} = \hat{\boldsymbol{z}}\Gamma_0\left[\frac{1}{c^2}\left(\frac{4t^2}{\tau^4} - \frac{2}{\tau^2}\right) - \left(\frac{4r^2}{\sigma^4} - \frac{6}{\sigma^2}\right)\right] e^{-r^2/\sigma^2 - t^2/\tau^2} \tag{43}$$

The physical current density $\boldsymbol{J}$ and charge density $\rho$ required to generate this wave packet are obtained directly from $\boldsymbol{G}$**.**

Eq. (17) gives the current density as $\boldsymbol{J} = \varepsilon_0 \frac{\partial \boldsymbol{G}}{\partial t}$, which, upon differentiating Eq. (43) with respect to time, becomes:

$$\boldsymbol{J}(\boldsymbol{r},t) = \hat{\boldsymbol{z}}\varepsilon_0\Gamma_0 \frac{4t}{\tau^2}\left[\frac{3}{c^2\tau^2} - \frac{2t^2}{c^2\tau^4} + \frac{2r^2}{\sigma^4} - \frac{3}{\sigma^2}\right] e^{-r^2/\sigma^2 - t^2/\tau^2} \tag{44}$$

The charge density, per Eq. (16), is $\rho_\Gamma = -\varepsilon_0 \nabla \cdot \boldsymbol{G}$. Calculating the divergence gives:

$$\rho(\boldsymbol{r},t) = \varepsilon_0 z\Gamma_0\left[-\frac{8r^2}{\sigma^6} + \frac{20}{\sigma^4} + \frac{1}{c^2\sigma^2}\left(\frac{8t^2}{\tau^4} - \frac{4}{\tau^2}\right)\right] e^{-r^2/\sigma^2 - t^2/\tau^2} \tag{45}$$

Eqs. (44) and (45) provide the precise, self-consistent source distribution necessary to support a Gaussian wave packet in vacuum.

Beyond the sources, the complete electromagnetic configuration—potentials and fields—follows immediately from the derivatives of $\boldsymbol{\Gamma}$.

The scalar potential $\varphi = -\nabla \cdot \boldsymbol{\Gamma}$ (Eq. (12)) and vector potential $\boldsymbol{A} = \frac{1}{c^2}\frac{\partial \boldsymbol{\Gamma}}{\partial t}$ (Eq. (13)) are derived as:

$$\varphi(\boldsymbol{r},t) = \frac{2z}{\sigma^2}\Gamma_0 e^{-r^2/\sigma^2 - t^2/\tau^2} \tag{46}$$

$$\boldsymbol{A}(\boldsymbol{r},t) = -\hat{\boldsymbol{z}}\frac{2t}{c^2\tau^2}\Gamma_0 e^{-r^2/\sigma^2 - t^2/\tau^2} \tag{47}$$

The electric field $\boldsymbol{E}$ and magnetic field $\boldsymbol{B}$, are obtained via Eqs. (14) and (15). The magnetic field yields a toroidal topology:

$$\boldsymbol{B}(\boldsymbol{r},t) = \frac{4t}{c^2\tau^2\sigma^2}\boldsymbol{\Gamma_0}(y\hat{\boldsymbol{x}} - x\hat{\boldsymbol{y}})e^{-r^2/\sigma^2 - t^2/\tau^2} \tag{48}$$

The electric field exhibits a dominant longitudinal component modified by transverse variations:

$$\boldsymbol{E}(\boldsymbol{r},t) = \boldsymbol{\Gamma_0}\left[\hat{\boldsymbol{z}}\left(\frac{2}{c^2\tau^2}\left(1 - \frac{2t^2}{\tau^2}\right) - \frac{2}{\sigma^2}\left(1 - \frac{2z^2}{\sigma^2}\right)\right) - \boldsymbol{r}_\perp\frac{4z}{\sigma^4}\right]e^{-r^2/\sigma^2 - t^2/\tau^2} \tag{49}$$

where $\boldsymbol{r}_\perp = x\hat{\boldsymbol{x}} + y\hat{\boldsymbol{y}}$ . These explicit expressions provide a complete analytical map of the electromagnetic configuration for the defined packet and, by construction, are consistent with Maxwell's equations, the potential–field relations, the Lorenz-gauge condition, and charge conservation.

**7.2. Comparison with the retarded-potential formulation**

In standard electrodynamics (e.g., [2, 3]), the potentials generated by a transient current distribution are given by retarded integrals. The vector potential, for example, is

$$\boldsymbol{A}(\boldsymbol{x},t) = \frac{\mu_0}{4\pi}\int_{\mathbb{R}^3}\frac{\boldsymbol{J}(\boldsymbol{x}', t - |\boldsymbol{x} - \boldsymbol{x}'|/c)}{|\boldsymbol{x} - \boldsymbol{x}'|}d^3x'. \tag{50}$$

For sources localized in both space and time, the retardation condition $t' = t - |\boldsymbol{x} - \boldsymbol{x}'|/c$ couples the spatial integration variables to the time dependence in a nonseparable way. For the Gaussian wave packet of Sec. 7.1, for instance, the temporal envelope is evaluated at the retarded time $t - |\boldsymbol{x} - \boldsymbol{x}'|/c$, yielding a factor of the form $exp[-(t - |\boldsymbol{x} - \boldsymbol{x}'|/c)^2/\tau^2]$, which complicates direct spacetime-domain evaluation of Eq. (50). As a result, explicit closed-form evaluation and simplification of the retarded integral for $\mathbf{A}(\mathbf{x},t)$ becomes analytically intractable in many cases. Practical treatments therefore often employ spectral methods, special-function representations, or numerical quadrature.

Accordingly, exact localized wave-packet fields are often constructed indirectly via spectral decompositions or other nonlocal representations. In the localized-wave literature (e.g., focus wave modes [8] and directed-energy pulse constructions [9]), the construction typically starts from special scalar wave solutions and then generates Maxwell fields via Hertz-potential-type mappings. In ultrafast and beam optics, related constructions are also developed via the complex-source-point method and subsequently interpreted in terms of physically realizable sources [10]. While these approaches are fully consistent in principle, they typically shift the complexity into the choice of special waveforms and nonlocal representations, which can be comparably challenging to manage analytically. As a result, compact real-spacetime expressions (together with an explicit supporting $\rho, \mathbf{J}$) are not always readily available for transient localized packets.

By contrast, the Γ-based route turns the inverse-source task into a purely local real-spacetime procedure: once $\boldsymbol{\Gamma}(\mathbf{x}, t)$ is prescribed with adequate regularity, the fields and the supporting sources follow from a fixed differential recipe ($\Box\boldsymbol{\Gamma} \rightarrow \boldsymbol{G} \rightarrow \rho, \boldsymbol{J}$), together with $\phi, \boldsymbol{A}, \boldsymbol{E}, \boldsymbol{B}$. In other words, the Γ-representation shifts the technical burden from explicit evaluation of retarded convolution integrals to local differentiation in real spacetime, making the construction substantially more tractable for transient localized packets. It therefore provides a more direct and practical route to explicit closed-form expressions for transient localized wave packets.

## 8. Summary and Outlook

In this work, we introduce the Γ-representation as a unified, constructive framework for describing electromagnetic configurations, in which fields, potentials, and sources are systematically reconstructed from a single vector wavefield. This extends the classical concept of Hertz potentials by generalizing the representation to encompass all sufficiently regular electromagnetic configurations, enabling the quantities and their interrelations—including Maxwell's equations—to be encoded in a single vector wavefield. The mapping between electromagnetic configurations and the vector wavefields reveals layered symmetries and a linear algebraic structure, enabling systematic tools for transformation and classification. The framework remains compatible with classical results while offering new viewpoints on solution construction and organization.

While electromagnetic fields and sources are traditionally viewed as the only physically observable quantities, phenomena such as the Aharonov–Bohm effect suggest that the electromagnetic potentials—scalar potential $\varphi$ and vector potential $\boldsymbol{A}$—may themselves possess physical significance beyond their role as mathematical intermediaries. In this context, the Γ-potential extends the classical three-tiered hierarchy (sources, fields, and the electromagnetic potentials) by adding a fourth layer that generates the electromagnetic potentials themselves. Should the electromagnetic potentials indeed have ontological status—i.e., possess physical reality—the Γ-potential may then serve as an auxiliary structure for them, much as the scalar and vector potentials underlie the observable fields, offering a fresh perspective on the foundations of electromagnetic theory.

Although the present work adopts a non-covariant (3+1) formulation to highlight the structural derivations and physical interpretation, the Γ-representation admits a covariant four-dimensional embedding. In such a setting, the Γ-potential may be reinterpreted as a spacetime vector field satisfying a covariant wave equation, potentially allowing a reformulation within relativistic field theory. Its transformation properties, relation to four-potentials, and compatibility with Lorentz invariance may merit further investigation.

**Acknowledgments.** The author is grateful to Markus Penz for insightful comments on regularity—particularly for noting that the regularity assumption in an earlier draft was insufficient—and for encouraging the "total-source → Γ-potential" construction in the media case. His suggestions substantially clarified the presentation.

**DATA AVAILABILITY** — No datasets were generated or analyzed in this study. All mathematical derivations needed to verify the results are contained in the manuscript.

**CONFLICT OF INTEREST** — The author declares no conflict of interest.

**AUTHOR CONTRIBUTIONS** — Conceptualization, Methodology, Formal analysis, Writing—original draft, Writing—review & editing: Ting Yi.

**ETHICS APPROVAL** — Not applicable. This theoretical work involves no human or animal subjects.